\RequirePackage{fix-cm}
\documentclass[smallextended]{svjour3}
\smartqed
\usepackage{graphicx}
\usepackage{amsmath}
\usepackage{amssymb}
\usepackage{graphicx}
\newcommand{\ket}[1]{|#1\rangle}
\newcommand{\bra}[1]{\langle #1|}
\newcommand{\Tr}{\mathrm{Tr}}

\newcommand{\abs}[1]{\lvert #1\rvert}
\newcommand{\ip}[2]{\langle #1|#2\rangle}
\journalname{Quantum Information Processing}

\begin{document}

\title{A new coherence measure based on fidelity}
\author{C. L. Liu\textsuperscript{1}\and Da-Jian Zhang\textsuperscript{1,2}\and Xiao-Dong Yu\textsuperscript{1}\and Qi-Ming Ding\textsuperscript{1}\and Longjiang Liu\textsuperscript{3}
}
\authorrunning{C. L. Liu et al.}

\institute{Longjiang Liu \\\email{wlliulongjiang@aliyun.com} \\
~\\
\textsuperscript{1} Department of Physics, Shandong University, Jinan, 250100, China\\
\textsuperscript{2} School of Physics and Electronics, Shandong Normal University, Jinan, 250014, China\\
\textsuperscript{3} College of Science, Henan University of Technology, Zhengzhou, 450001, China}

\date{Received: date / Accepted: date}

\maketitle

\begin{abstract}
Quantifying coherence is an essential endeavor for both quantum foundations and quantum technologies. In this paper, we put forward a quantitative  measure of coherence by following the axiomatic definition of coherence measures introduced in [T. Baumgratz, M. Cramer, and M. B. Plenio, Phys. Rev. Lett. \textbf{113}, 140401 (2014)]. Our measure is based on fidelity and analytically computable for arbitrary states of a qubit. As one of its applications, we show that our measure can be used to examine whether a pure qubit state can be transformed into another pure or mixed qubit state only by incoherent operations.
\keywords{Quantifying coherence \and Coherence measure \and Fidelity\and Qubit states}

\end{abstract}

\section{Introduction}
\label{intro}
Coherence is a fundamental aspect of quantum physics, encapsulating the defining features of the theory,  from  the  superposition  principle  to  quantum  correlations. It is an essential component in quantum information processing \cite{Nielsen}, and plays a central role in emergent fields, such as quantum metrology \cite{Lloyd,Dobrzanski2014}, nanoscale thermodynamics \cite{Aberg,Lostaglio}, and quantum biology \cite{Sarovar,Lloyd1,Huelga,Lambert}. Although the theory of coherence is historically well developed in quantum optics \cite{Glauber,Sudarshan,Mandel}, it is only in recent years that the quantification of coherence has attracted a growing interest due to the development of quantum information science \cite{Gour,Marvian,Levi,Baumgratz,Aberg1,Yu}.

By adopting the viewpoint of coherence as a physical resource, Baumgratz \textit{et al.} proposed a seminal framework for quantifying coherence \cite{Baumgratz}. In this framework, a functional is defined to be a legitimate coherence measure if it fulfills four conditions, namely, the coherence being zero (positive) for incoherent states (all other states), the monotonicity of coherence under incoherent operations, the monotonicity of coherence under selective measurements on average, and the nonincreasing of coherence under mixing of quantum states. By following the rigorous framework, a number of coherence measures, such as the $l_1$ norm of coherence \cite{Baumgratz}, the relative entropy of coherence \cite{Baumgratz}, the distillable coherence \cite{Winter,Yuan}, the coherence of formation \cite{Aberg1,Winter,Yuan}, the robustness of coherence \cite{Napoli}, the coherence measures based on entanglement \cite{Streltsov}, and the coherence concurrence \cite{Qi,Du}, have been proposed. These measures have been widely used to study various topics related to coherence, such as the freezing phenomenon of coherence \cite{Bromley,Yu1,Zhang}, the relation between coherence and other quantum resources \cite{Streltsov,Killoran,Xi,Yao,Ma},  the complementarity between coherence and mixedness \cite{Cheng,Singh1}, the relations between coherence and path information \cite{Bera,Bagan}, the distribution of quantum coherence in multipartite systems \cite{Radhakrishnan}, the phenomenon of coherence sudden death \cite{Bu}, and the ordering states with coherence measures \cite{Liu,Zhangf}.

In this paper, we put forward a quantitative measure of coherence by following the axiomatic definition of coherence measures introduced in Ref. \cite{Baumgratz}. Our measure is based on fidelity and analytically computable for arbitrary states of a qubit. As one of its applications, we show that our measure can be used to examine whether a pure qubit state can be transformed into another pure or mixed qubit state only by incoherent operations.

This paper is organized as follows. In Sec. 2, we review the framework for  quantifying coherence introduced in Ref. \cite{Baumgratz}. In Sec. 3, we give the definition of our measure and show that it fulfills the conditions proposed in Ref. \cite{Baumgratz}. In Sec. 4, we show that our measure is analytically computable for arbitrary states of a qubit. In Sec. 5, we show that our measure can be used to examine whether a pure qubit state can be transformed into another pure or mixed qubit state only by incoherent operations. Section 6 is our conclusion.
\section{Framework for quantifying coherence}
\label{sec:1}
We first specify some notions introduced in the framework for quantifying coherence, such as incoherent states, incoherent operations, and coherence measures \cite{Baumgratz}.

Let us consider a quantum system equipped with a $d$-dimensional Hilbert space. Coherence of a state is measured with respect to a particular reference basis, whose choice is dictated by the physical scenario under consideration. The particular basis is denoted as $\{\ket{i}, ~i=0,1,\cdot\cdot\cdot,d-1\}$. A state is called an incoherent state if its density operator is diagonal in the basis, and the set of all incoherent states is denoted by $\mathcal {I}$. It follows that a density operator $\delta$ belonging to $\mathcal {I}$ is of the form
$\delta=\sum^{d-1}_{i=0}\delta_{i}\ket{i}\bra{i}$.
All other states, which cannot be written as diagonal matrices in the basis, are called coherent states. Hereafter, we use $\rho$ to represent a general state, a coherent state or an incoherent state, and use $\delta$ specially to denote an incoherent state.

A completely positive trace-preserving map, $\Lambda(\rho)=\sum_nK_n \rho K_n^\dag$, is said to be an incoherent completely positive trace-preserving (ICPTP) map or an incoherent operation, if the Kraus operators $K_n$ satisfy not only $\sum_nK_n^\dag K_n=I$  but also $ K_n \mathcal {I} K_n^\dag \subseteq \mathcal {I} $, i.e., each $K_n$ maps an incoherent state to an incoherent state.

A functional $C$ can be taken as a legitimate measure of coherence if it satisfies the following four conditions \cite{Baumgratz}:
\begin{itemize}
\item[(C1)] $C(\rho)\geq0$, and $C(\rho)=0$  if and only if $\rho\in \mathcal {I} $;
\item[(C2)] Monotonicity under incoherent completely positive
and trace preserving maps, i.e., $C(\rho)\geq C(\Lambda_{\textrm{ICPTP}}(\rho))$ for all ICPTP maps $\Lambda_{\textrm{ICPTP}}$;
\item[(C3)] Monotonicity under selective measurements on average, i.e., $C(\rho)\geq\sum_n p_nC\left(\rho_n\right)$, where $p_n=\Tr(K_n \rho K_n^\dagger)$ and $ \rho_n=K_n \rho K_n^\dagger/p_n$, with $K_n$ satisfying $\sum_n K_n^\dagger K_n=I$ and $K_n\mathcal{I}K_n^\dagger\subset\mathcal{I}$;
\item[(C4)] Nonincreasing under mixing of quantum states (convexity), i.e., $C(\sum_n p_n\rho_n)\leq\sum_n p_nC(\rho_n)$ for any set of states $\{\rho_n\}$ and any $p_n\geq0$ with $\sum_n p_n=1$.
\end{itemize}
Note that conditions (C3) and (C4) automatically imply condition (C2).

Among various coherence measures, the $l_1$-norm quantifies coherence in an intuitive way. It can be expressed as
\begin{eqnarray}
C_{l_1}(\rho)=\sum_{i\neq j}|\rho_{ij}|,\label{L}
\end{eqnarray}
where $\rho_{ij}$ are entries of $\rho$ in the basis.

\section{Coherence measure based on fidelity}
\label{sec:2}

In this section, we give the definition of our measure and then prove that our measure fulfills the four conditions introduced in Ref. \cite{Baumgratz}.

The definition is based on convex-roof construction. We define our measure for a pure state as
\begin{equation}
C_F(\ket{\varphi})= \min_{\delta\in\mathcal{I}}\sqrt{1-F(\ket{\varphi},\delta)}, \label{geo-con-pure}
\end{equation}
where $F({\rho,\sigma})=[\Tr{(\sqrt{\sqrt{\rho}\sigma\sqrt{\rho}})}]^2$ is the Uhlmann fidelity. We then extend our definition to the general case via convex-roof construction,
\begin{equation}
C_F(\rho)=\min_{\{p_n,\ket{\varphi_n}\}}\sum_{n}p_nC_F(\ket{\varphi_n}), \label{geo-mixed-ana}
\end{equation}
where the minimum is taken over all the ensembles $\{p_n,\ket{\varphi_n}\}$ realizing $\rho$, i.e.,  $\rho=\sum_{n}p_n\ket{\varphi_n}\bra{\varphi_n}$.

By definition, for a pure state $\ket{\varphi}=\sum_{i=0}^{d-1} c_i\ket{i}$, where $c_i$ are complex numbers satisfying $\sum_{i=0}^{d-1}\abs{c_i}^2=1$, there is
\begin{equation}
\begin{aligned}
C_F(\ket{\varphi})&=\min_{\delta\in\mathcal{I}}\sqrt{1-F(\ket{\varphi},\delta)}\\&=\min_{\delta\in\mathcal{I}}\sqrt{1-\bra{\varphi}\delta\ket{\varphi}}
\\&=\min_{\delta\in\mathcal{I}}\sqrt{1-\sum_i\delta_i\abs{c_i}^2}\\&=\sqrt{1-\abs{c_i}^2_{\max}},
\end{aligned}
\end{equation}
where $\abs{c_i}_{\max}=\max\{\abs{c_0},\abs{c_1},\cdots, \abs{c_{d-1}}\}$. That is
\begin{equation}
C_F(\ket{\varphi})=\sqrt{1-\abs{c_i}^2_{\max}}. \label{geo-pure-ana}
\end{equation}
Equation (\ref{geo-pure-ana}) implies that $C_F(\ket{\varphi})=0$ if and only if $\ket{\varphi}$ is a pure incoherent state.

With the above knowledge, we now prove that the functional defined by Eq. (\ref{geo-mixed-ana}) with Eq. (\ref{geo-con-pure}) satisfies conditions (C1)-(C4) and hence is a legitimate coherence measure.

First, we show that the functional defined by Eq. (\ref{geo-mixed-ana}) with Eq. (\ref{geo-con-pure}) satisfies condition (C1). By definition, there is $C_F(\rho)\geq 0$. Since an incoherent state admits a decomposition of the form $\delta=\sum_{i=0}^{d-1}\delta_i\ket{i}\bra{i}$, we have $C_F(\delta)\leq \sum_{i=0}^{d-1} p_iC_F(\ket{i})=0$. Hence, $C_F(\delta)=0$ for an incoherent state $\delta$. Conversely, suppose that $C_F(\rho)=0$ for a state $\rho$. Then, there exists an ensemble $\{p_n,\ket{\varphi_n}\}$ of $\rho$ such that $\sum_np_nC_F(\ket{\varphi_n})=0$, which further leads to $C_F(\ket{\varphi_n})=0$ for all $n$. It follows that each $\ket{\varphi_n}$ is an incoherent state, and so is $\rho$.

Second, we prove that the functional defined by Eq. (\ref{geo-mixed-ana}) with Eq. (\ref{geo-con-pure}) satisfies condition (C4). Let $\{\rho_n\}$ be a set of states and $p_n$ be probabilities, and let $\{q_m^n,\ket{\varphi_m^n}\}$ be the ensemble of $\rho_n$ achieving the minimum in the definition of $C_F(\rho_n)$. We then have
\begin{equation}
\begin{aligned}
\sum_n p_nC_F(\rho_n)&=\sum_n p_n\sum_mq_m^nC_F(\ket{\varphi_m^n})\\&\geq C_F\left(\sum_{n,m}p_nq_m^n\ket{\varphi_m^n}\bra{\varphi_m^n}\right)\\
&=C_F\left(\sum_np_n\rho_n\right),\label{C4}
\end{aligned}
\end{equation}
where the second inequality follows from the definition of $C_F$.

Third, we prove that the functional defined by Eq. (\ref{geo-mixed-ana}) with Eq. (\ref{geo-con-pure}) satisfies condition (C3). We first consider the pure-state case, in which we need to show that the inequality
\begin{equation}
C_F(\ket{\varphi})\geq\sum_{n}p_n C_F\left(\frac{1}{\sqrt{p_n}}K_n\ket{\varphi}\right), \label{c3-1}
\end{equation}
holds for an arbitrary pure state $\ket{\varphi}$, where $p_n=\Tr(K_n\ket{\varphi}\bra{\varphi}K_n^\dagger)$.

Note that a pure state and a Kraus operator can be expressed, without loss of generality, as $\ket{\varphi}=\sum_{i=0}^{d-1}c_i\ket{i}$ and $K_n=\sum_i\ket{i}\bra{\phi_i^n}$, repectively, where $\ket{\phi_i^n}$ are unnormalized states. Since $K_n$ belongs to an incoherent operation, $K_n$ maps an incoherent pure state to an incoherent pure state, and hence at most one of the terms $\ip{\phi_0^n}{i}$, $\dots$, $\ip{\phi_{d-1}^n}{i}$ is nonzero for all $i=0,\dots,d-1$.

We assume that $\abs{c_{i_0}}=\max\{\abs{c_0},\dots,\abs{c_{d-1}}\}$, where $0\leq i_0\leq d-1$ is a fixed integer. Using Eq. (\ref{geo-pure-ana}), we have
\begin{eqnarray}\label{c3_2}
C_F(\ket{\varphi})=\sqrt{1-\abs{c_{i_0}}^2}.
\end{eqnarray}
On the other hand, we have $K_n\ket{\varphi}=\sum_{i=0}^{d-1}\ip{\phi_i^n}{\varphi}\ket{i}$, which further leads to that
\begin{eqnarray}\label{c3_3}
\sum_np_nC_F\left(\frac{1}{\sqrt{p_n}}K_n\ket{\varphi}\right)=\sum_np_n\sqrt{1-\frac{\abs{\ip{\phi_i^n}{\varphi}}_{\max}^2}{p_n}},
\end{eqnarray}
where $\abs{\ip{\phi_i^n}{\varphi}}_{\max}=\max\{\abs{\ip{\phi_0^n}{\varphi}},\dots,\abs{\ip{\phi_{d-1}^n}{\varphi}}\}$.

Note that at most one of the terms $\ip{\phi_0^n}{i_0}$, $\dots$, $\ip{\phi_{d-1}^n}{i_0}$ is nonzero. Suppose that the nonzero term is the $i_n$-th one, i.e., $\ip{\phi_{i_n}^n}{i_0}\neq 0$ and $\ip{\phi_{i}^n}{i_0}= 0$ for all $i\neq i_n$. It follows that
\begin{equation}
\begin{aligned}\label{c3_4}
\bra{i_0}\left(\sum_n\ket{\phi_{i_n}^n}\bra{\phi_{i_n}^n}\right)\ket{i_0}
&=\bra{i_0}\left(\sum_{i,n}\ket{\phi_{i}^n}\bra{\phi_{i}^n}\right)\ket{i_0}\\
&=\bra{i_0}\left(\sum_n K_n^\dagger K_n\right)\ket{i_0}=1.
\end{aligned}
\end{equation}
Noting that $0\leq\sum_n\ket{\phi_{i_n}^n}\bra{\phi_{i_n}^n}\leq I$, which means that $\sum_n\ket{\phi_{i_n}^n}\bra{\phi_{i_n}^n}$ is a positive semi-definite operator with the largest eigenvalue being 1, we deduce from Eq. (\ref{c3_4}) that $\ket{i_0}$ is a eigenvector of $\sum_n\ket{\phi_{i_n}^n}\bra{\phi_{i_n}^n}$ corresponding to the eigenvalue 1. As an immediate consequence, we have
\begin{eqnarray}\label{c3_5}
\sum_n\ket{\phi_{i_n}^n}\bra{\phi_{i_n}^n}\geq\ket{i_0}\bra{i_0}.
\end{eqnarray}

Using Eqs. (\ref{c3_2}), (\ref{c3_3}), and (\ref{c3_5}), we have
\begin{eqnarray}\label{c3_6}
\sum_np_nC_F\left(\frac{1}{\sqrt{p_n}}K_n\ket{\varphi}\right)&=&\sum_np_n\sqrt{1-\frac{\abs{\ip{\phi_i^n}{\varphi}}_{\max}^2}{p_n}}\nonumber\\
&\leq&\sqrt{1-\sum_n\abs{\ip{\phi_i^n}{\varphi}}_{\max}^2}\nonumber\\
&\leq&\sqrt{1-\sum_n\abs{\ip{\phi_{i_n}^n}{\varphi}}^2}\nonumber\\
&=&\sqrt{1-\bra{\varphi}\left(\sum_n\ket{\phi_{i_n}^n}\bra{\phi_{i_n}^n}\right)\ket{\varphi}}\nonumber\\
&\leq&\sqrt{1-\ip{\varphi}{i_0}\ip{i_0}{\varphi}}\nonumber\\
&=&\sqrt{1-\abs{c_{i_0}}^2}=C_F(\ket{\varphi}),
\end{eqnarray}
where we have used the concavity of the function $\sqrt{x}$, namely, $\sum_np_n\sqrt{x_n}\leq\sqrt{\sum_np_nx_n}$, with $p_n$ being probabilities and $x_n$ being non-negative numbers.
Hence, we have proved Eq. (\ref{c3-1}).

We now consider the general case. By definition, for a generic state $\rho$, there exists an ensemble, denoted by $\{q_m,\ket{\varphi_m}\}$, such that $C_F(\rho)=\sum_mq_mC_F(\ket{\varphi_m})$.
With the aid of Eq. (\ref{c3-1}) and using the convexity of $C_F$, we have
\begin{equation}\label{c3_7}
\begin{aligned}
C_F(\rho)&=\sum_mq_mC_F(\ket{\varphi_m})\\&\geq\sum_mq_m\sum_n\Tr{(K_n\ket{\varphi_m}\bra{\varphi_m}K_n^\dagger)}C_F(\frac{K_n\ket{\varphi_m}\bra{\varphi_m}K_n^\dagger}{\Tr{(K_n\ket{\varphi_m}\bra{\varphi_m}K_n^\dagger)}})\\&=\sum_n\Tr{(K_n\rho K_n^\dagger)}\sum_m\frac{q_m\Tr{(K_n\ket{\varphi_m}\bra{\varphi_m}K_n^\dagger)}}{\Tr{(K_n\rho K_n^\dagger)}}C_F(\frac{K_n\ket{\varphi_m}\bra{\varphi_m}K_n^\dagger}{\Tr{(K_n\ket{\varphi_m}\bra{\varphi_m}K_n^\dagger)}})\\&\geq\sum_n\Tr{(K_n\rho K_n^\dagger)}C_F({\frac{K_n\rho K_n^\dagger}{\Tr{(K_n\rho K_n^\dagger)}}}).
\end{aligned}
\end{equation}
Equation (\ref{c3_7}) shows that the functional defined by Eq. (\ref{geo-mixed-ana}) with Eq. (\ref{geo-con-pure}) satisfies condition (C3).

Since (C3) and (C4) imply (C2), we obtain that the functional defined by Eq. (\ref{geo-mixed-ana}) with Eq. (\ref{geo-con-pure}) satisfies condition (C2), too, thus completing the proof.

\section{Analytic expression for arbitrary single-qubit states} \label{IV.5}

After proving that $C_F$ is a legitimate coherence measure obeying (C1)-(C4), we show that $C_F$ is analytically computable for arbitrary states of a qubit. We present our result as the following proposition.

\textit{Proposition 1.} For an arbitrary single-qubit state $\rho$, $C_F(\rho)$ admits the following expression,
\begin{eqnarray}\label{ana_exp}
C_F(\rho)=\sqrt{\frac{1-\sqrt{1-4\abs{\rho_{01}}^2}}2},
\end{eqnarray}
where $\rho_{01}$ is the off-diagonal element of $\rho$ with respect to the reference basis.

It is worth noting that $C_F$ in this case is a simple monotonic function of the $l_1$-norm of coherence, $C_{l_1}(\rho)=2\abs{\rho_{01}}$.

We prove our result step by step in the following.

First, we introduce an auxiliary state $\tilde{\rho}$, defined as $\tilde{\rho}_{00}=\rho_{00}$, $\tilde{\rho}_{01}=\abs{\rho_{01}}$, $\tilde{\rho}_{10}=\abs{\rho_{10}}$, and $\tilde{\rho}_{11}=\rho_{11}$, and  show that $C_F(\rho)=C_F(\tilde{\rho})$.
This can be easily proved by resorting to the incoherent unitary operator $U:=\textrm{diag}(1,\exp[i\arg(\rho_{01})])$. Indeed, since $U\rho U^\dagger=\tilde{\rho}$ and $\rho=U^\dagger\tilde{\rho}U$, the equality $C_F(\rho)=C_F(\tilde{\rho})$ follows immediately from condition (C2).

Second, we show that there exists an ensemble $\{\tilde{p}_n,\ket{\tilde{\varphi}_n}\}$ of $\tilde{\rho}$ such that $C_F(\ket{\tilde{\varphi}_n})=f(\abs{\rho_{01}})$ for each $n$, where $f(x):=\sqrt{(1-\sqrt{1-4x^2})/2}$. To this end, we introduce the pure states defined by
\begin{eqnarray}
\ket{\tilde{\varphi}_1}=\sqrt{q}\ket{0}+\sqrt{1-q}\ket{1},
\end{eqnarray}
and
\begin{eqnarray}
\ket{\tilde{\varphi}_2}=\sqrt{1-q}\ket{0}+\sqrt{q}\ket{1},
\end{eqnarray}
respectively, where $q$ is an non-negative number satisfying $\sqrt{q(1-q)}=\abs{\rho_{01}}$. Since $\sqrt{q(1-q)}=\abs{\rho_{01}}\leq\sqrt{\rho_{00}\rho_{11}}=\sqrt{\rho_{00}(1-\rho_{00})}$, we have that $\rho_{00}$ lies between $q$ and $(1-q)$. Hence, there exists a number $0\leq \tilde{p}_1\leq 1$ such that $\rho_{00}=\tilde{p}_1q+\tilde{p}_2(1-q)$, where $\tilde{p}_2=1-\tilde{p}_1$. Direct calculations show that $\tilde{\rho}=\tilde{p}_1\ket{\tilde{\varphi}_1}\bra{\tilde{\varphi}_1}+\tilde{p}_2\ket{\tilde{\varphi}_2}\bra{\tilde{\varphi}_2}$ and $C_F(\ket{\tilde{\varphi}_1})=C_F(\ket{\tilde{\varphi}_2})=f(\abs{\rho_{01}})$. Thus, we arrive at the desired ensemble of $\tilde{\rho}$.

Third, with the aid of the auxiliary state and the ensemble, we are ready to prove Eq. (\ref{ana_exp}). Note that for a pure qubit state $\ket{\varphi}$, $C_F(\ket{\varphi})=f(\abs{x_\varphi})$, where $x_\varphi$ denotes the off-diagonal element of $\ket{\varphi}\bra{\varphi}$, and also note that $f(x)$ is a convex and monotonically increasing function. For an arbitrary ensemble $\{p_n,\ket{\varphi_n}\}$ of $\tilde{\rho}$, we have
\begin{eqnarray}\label{ana-1}
\sum_np_nC_F(\ket{\varphi_n})&=&\sum_np_nf(\abs{x_{\varphi_n}})\nonumber\\
&\geq&f(\sum_np_n\abs{x_{\varphi_n}})\nonumber\\
&\geq&f(\abs{\sum_np_nx_{\varphi_n}})\nonumber\\
&=&f(\abs{\rho_{01}})\nonumber\\
&=&\sum_n\tilde{p}_nC_F(\ket{\tilde{\varphi}_n}).
\end{eqnarray}
Equation (\ref{ana-1}) shows that $C_F(\tilde{\rho})=\sum_n\tilde{p}_nC_F(\ket{\tilde{\varphi}_n})=f(\abs{\rho_{01}})$. Hence, there is $C_F(\rho)=f(\abs{\rho_{01}})$. This completes the proof of Proposition 1.

\section{Application} \label{V}

In the resource theory of coherence, one important issue is to identify conditions under which one state can be transformed into another via incoherent operations \cite{Baumgratz}. In this section, we will show that our measure can be used to determine whether a pure qubit state can be transformed to another pure or mixed qubit state under incoherent operations. We present our result as the following proposition.

\emph{Proposition 2.} A pure qubit state $\ket{\phi}$ can be transformed into another pure or mixed qubit state $\rho$ via incoherent operations if and only if
$C_F(\ket{\phi})\geq C_F(\rho).$

Proposition 2 is easy to be proved. The ``only if'' part follows directly from the fact that the measure $C_F$ satisfies condition (C2). Therefore, we only need to prove the ``if'' part. To this end, we simply assume that $\ket{\phi}$ is of the form
$\ket{\phi}=\sqrt{p}\ket{0}+\sqrt{1-p}\ket{1}$; otherwise, $\ket{\phi}$ can be transformed into this form via an incoherent unitary operator.
In the proof of Proposition 1, we have shown that $\rho$ can be transformed via an incoherent unitary operator into $\tilde{\rho}$, which is defined as
$\tilde{\rho}=\tilde{p}_1\ket{\tilde{\varphi}_1}\bra{\tilde{\varphi}_1}+\tilde{p}_2\ket{\tilde{\varphi}_2}\bra{\tilde{\varphi}_2}$,
where $\ket{\tilde{\varphi}_1}=\sqrt{q}\ket{0}+\sqrt{1-q}\ket{1}$ and $\ket{\tilde{\varphi}_2}=\sqrt{1-q}\ket{0}+\sqrt{q}\ket{1}$. Hence, we can simply assume that $\rho$ is of the form $\rho=\tilde{p}_1\ket{\tilde{\varphi}_1}\bra{\tilde{\varphi}_1}+\tilde{p}_2\ket{\tilde{\varphi}_2}\bra{\tilde{\varphi}_2}$.
For ease of notation, we also assume that $p\geq 1/2$ and $q\geq 1/2$. In the condition of $C_F(\ket{\phi})\geq C_F(\rho)$, we have that $p\leq q$. With the above knowledge, we now construct the ICPTP map transforming $\ket{\phi}$ into $\rho$, which is defined as $\Lambda(\rho)=\sum_{n=1}^4K_n\rho K_n^\dagger$, with
\begin{equation}\label{pro2}
\begin{aligned}
K_1&=\sqrt{\frac{\tilde{p}_1(p+q-1)}{2q-1}}
\begin{bmatrix}
\sqrt{\frac{q}{p}} & 0\\
0 & \sqrt{\frac{1-q}{1-p}}
\end{bmatrix},\\
K_2&=\sqrt{\frac{\tilde{p}_1(q-p)}{2q-1}}
\begin{bmatrix}
0 & \sqrt{\frac{q}{1-p}}\\
\sqrt{\frac{1-q}{p}} & 0
\end{bmatrix},\\
K_3&=\sqrt{\frac{\tilde{p}_2(p+q-1)}{2q-1}}
\begin{bmatrix}
0 & \sqrt{\frac{1-q}{1-p}}\\
\sqrt{\frac{q}{p}} & 0
\end{bmatrix},\\
K_4&=\sqrt{\frac{\tilde{p}_2(q-p)}{2q-1}}
\begin{bmatrix}
\sqrt{\frac{1-q}{p}} & 0\\
0 & \sqrt{\frac{q}{1-p}}
\end{bmatrix}.
\end{aligned}
\end{equation}
Direct calculations show that $\Lambda$ is ICPTP and $\Lambda(\ket{\phi}\bra{\phi})=\rho$. This completes the proof of Proposition 2 \cite{note}.

\section{Conclusion}
In conclusion, we have put forward a quantitative measure of coherence by following the axiomatic definition of coherence measures introduced in Ref. \cite{Baumgratz}. Our measure is based on fidelity and analytically computable for arbitrary states of a qubit. As one of its applications, we have shown that our measure provides a criterion for determining whether a pure qubit state can be transformed into another pure or mixed qubit state only by incoherent operations.

\textbf{\emph{Note added.}}
We notice that a new paper, Ref. \cite{Zhu}, has recently provided a general discussion on coherence measures obtained by convex roof. By using the result in that paper, our proof of $C_F$ satisfying (C1)-(C4) can be further reduced.

\begin{acknowledgements}
This work was supported by the National Natural Science Foundation of China through Grant No. 11575101 and No. 11547113. D.J.Z. acknowledges support from the China Postdoctoral Science Foundation under Grant No. 2016M592173.
\end{acknowledgements}


\begin{thebibliography}{99}
\bibitem{note} In the proof, we have assumed that $p\neq 1$ and $q\neq 1/2$. It is easy to see that Proposition 2 is also valid for the case of $p=1$ or $q=1/2$.
\end{thebibliography}
\end{document}